\documentclass[hidelinks,12pt]{iopart}
\expandafter\let\csname equation*\endcsname\relax
\expandafter\let\csname endequation*\endcsname\relax\usepackage{amsmath}
\usepackage{amsfonts}
\usepackage{hyperref}
\usepackage{graphicx}
\usepackage{float}
\usepackage{subfigure}
\usepackage{ulem}
\usepackage{cite}

\begin{document}

\title[Helium transport on MAST]{Neoclassical and gyrokinetic analysis of time-dependent helium transport experiments on MAST}

\author{S. S. Henderson$^{1,2}$, L. Garzotti$^2$, F. J. Casson$^2$, 
        D. Dickinson$^2$, M. F. J. Fox$^{2,3}$, M. O'Mullane$^1$, A. Patel$^2$, 
	      C. M. Roach$^2$, H. P. Summers$^1$, M. Valovi\v{c}$^2$ 
	      and the MAST team}

\address{$^1$ Department of Physics SUPA, University of Strathclyde, Glasgow G4 ONG, UK}
\address{$^2$ EURATOM/CCFE Fusion Association, Culham Science Centre, Abingdon, Oxfordshire OX14 3DB, UK}
\address{$^3$ Rudolf Peierls Centre for Theoretical Physics, University of Oxford, Oxford OX1 3NP, UK}

\ead{stuart.henderson@ccfe.ac.uk}

\begin{abstract}
Time-dependent helium gas puff experiments have been performed on the Mega Amp\`{e}re Spherical Tokamak (MAST) during a two point plasma current scan in L-mode and a confinement scan at 900 kA. An evaluation of the He II ($n=4\rightarrow3$) spectrum line induced by charge exchange suggests anomalous rates of diffusion and inward convection in the outer regions of both L-mode plasmas. Similar rates of diffusion are found in the H-mode plasma, however these rates are consistent with neoclassical predictions. The anomalous inward pinch found in the core of L-mode plasmas is also not apparent in the H-mode core. Linear gyrokinetic simulations of one flux surface in L-mode using the \textsc{gs2} and \textsc{gkw} codes find that equilibrium flow shear is sufficient to stabilise ITG modes, consistent with BES observations, and suggest that collisionless TEMs may dominate the anomalous helium particle transport. A quasilinear estimate of the dimensionless peaking factor associated with TEMs is in good agreement with experiment. Collisionless TEMs are more stable in H-mode because the electron density gradient is flatter. The steepness of this gradient is therefore pivotal in determining the inward neoclassical particle pinch and the particle flux associated with TEM turbulence. 
\end{abstract}

\section{Introduction}\label{sec:intro}
Impurity transport is a subject of fundamental importance in plasma physics in general and in tokamak physics in particular. The behaviour of the various impurity species and the evolution of their concentration determines, among other things, the fuel dilution and the fusion reaction rate, the plasma radiation pattern and the local energy balance, the plasma effective charge, $Z_{eff}$, and resistivity and the neutral beam particle and power deposition profile. It is therefore important to develop both a sound experimental base and reliable models to interpret the experimental results and to predict the transport properties of impurities.

In conventional tokamaks impurity transport has been studied since the very early days of tokamak research (for a comprehensive review of
the subject see, for example, references \cite{Garbet2004,Dux2004,Guirlet2006a}). However, on spherical tokamaks (ST) this subject has been explored to a lesser extent due to the fact that the ST is a configuration developed in more recent times with respect to the conventional tokamak and that other subjects have been given higher priority so far. Experiments have been performed on the National Spherical Tokamak Experiment (NSTX) aimed at the characterisation of the transport properties of neon, lithium and carbon in H-mode \cite{Delgado-Aparicio2009, Scotti2013} and neon in L-mode \cite{Stutman2003}. Some work has also been done on the Mega Amp\`{e}re Spherical Tokamak (MAST) on the behaviour of tin \cite{Foster2007} and carbon \cite{McCone2010}. During the last two MAST experimental campaigns, further experiments have been performed to expand the experimental measurements of light impurity transport and to improve the quality of the measurement of the evolution of the impurity concentration.

The analysis presented in this paper indicates that helium transport is neoclassical in H-mode and in the L-mode core, whereas anomalous transport is the dominant mechanism in the outer radii of L-mode plasmas. These conclusions were reached by injecting short gas puffs of helium in to different plasma scenarios and measuring the subsequent evolution of the impurity concentration using charge exchange (CX) spectroscopy. Helium transport coefficients have been obtained in an interpretative way, by analysing the measured evolution of the helium particle flux and density gradient at each radii, and in a predictive way, using the \textsc{sanco} \cite{Lauro1994} impurity transport code to simulate the evolution of the helium density and then by fitting the helium diffusivity and convective velocity to minimise the difference between the measured and simulated profiles.
 
The aim of this paper is to compare the measured helium transport with the expected levels of neoclassical transport (induced by collisions with the main plasma ions) and anomalous transport (induced by the turbulence of the background plasma). For this analysis we have used three codes: \textsc{nclass} \cite{Houlberg1997} for neoclassical transport, and the gyrokinetic codes \textsc{gs2} \cite{Kotschenreuther1995,Dorland2000} and \textsc{gkw} \cite{Peeters2009} to give quasilinear estimates for the level of anomalous transport. The analysis performed with these codes has also allowed us to propose explanations of the differences between various plasma scenarios and to identify the main drives of neoclassical and anomalous transport.

The structure of the paper is the following: in section \ref{sec:experiment} we describe the experimental and diagnostic set-up and the parameter space covered by
the different plasma scenarios. In section \ref{sec:measurements}, we present the experimental measurements, describe in detail the analysis technique and give the resulting transport coefficients. Furthermore, we discuss these experimental results and interpret them in the context of our models for neoclassical and anomalous transport. Lastly, a summary of the results are given in section \ref{sec:conclusions}.

\section{Experiment Details}\label{sec:experiment}

\subsection{Plasma Scenarios}\label{sec:scenarios}
The reference scenario chosen for the impurity transport experiments described in this paper is an L-mode plasma with plasma current $I_p = 900$ kA, 
toroidal field $B_T = 0.55$ T, additional NBI heating power $P_{NBI} = 2.1$ MW, on-axis electron density $n_e = 3.5 \cdot 10^{19}$ m$^{-3}$, on-axis
 electron temperature $T_e = 1.0$ keV, major radius $R_0 = 0.83$ m, minor radius $a = 0.6$ m, elongation $\kappa = 1.93$ and triangularity 
$\delta =0.4$. The plasma operates with a double null divertor and is kept in L-mode by shifting the vertical position of the magnetic axis above the equatorial plane of the machine by $\sim5$ cm, a technique that has proved effective on MAST to prevent the L-H transition.

\begin{figure}[p]
	\centering
		\includegraphics[width=0.9\textwidth]{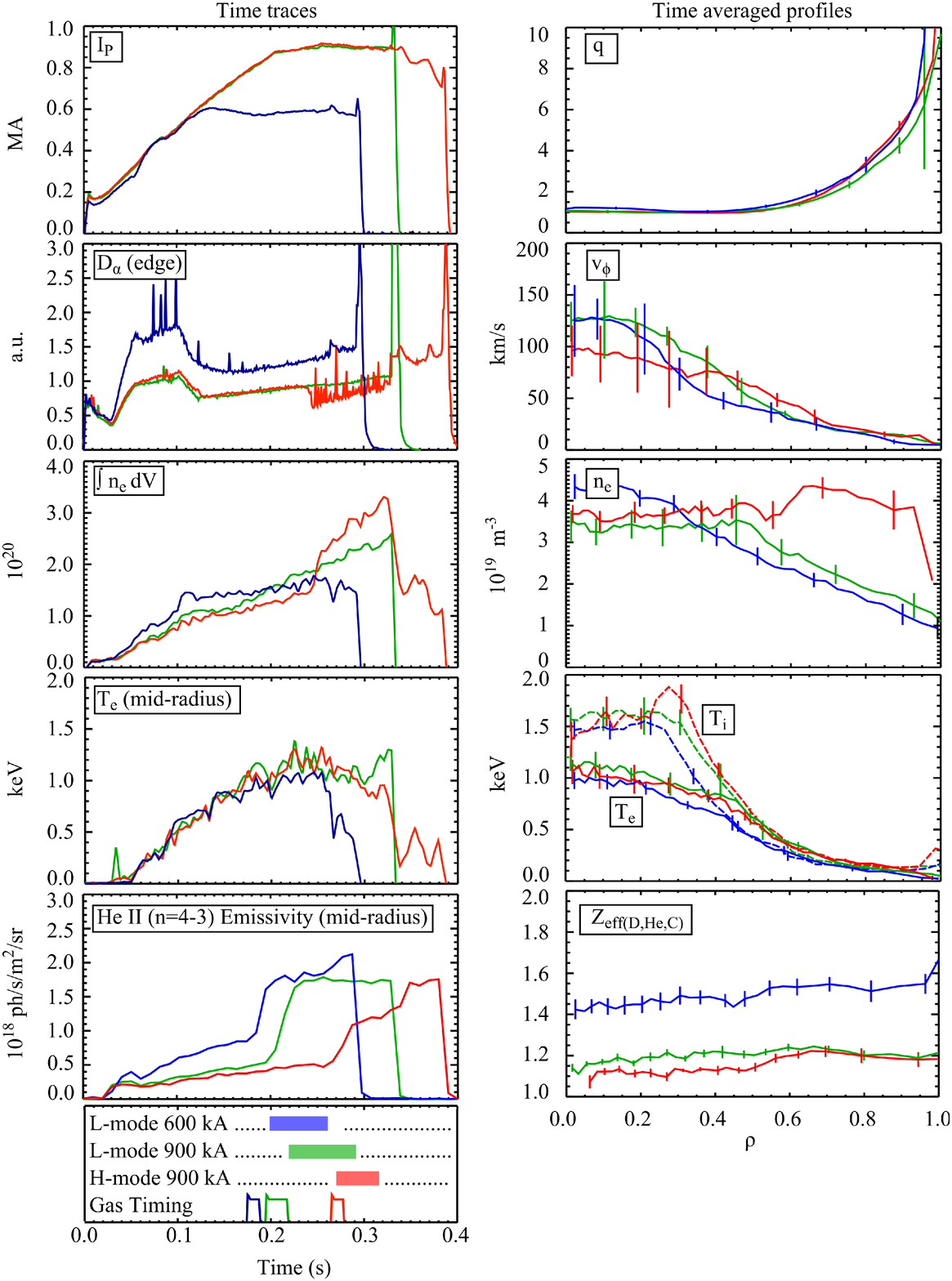}
	\caption{Time traces on the left and time averaged profiles on the right for shots \#29261 (green), \#29417 (blue) and \#29275 (red). Transport analysis and gas puff times are shown in the bottom panel. The spatial coordinate $\rho$ is the square root of the normalised toroidal flux.}
	\label{fig:trace}
\end{figure}

The transport analysis time window and the temporal plasma traces, including $I_p$, the edge $D_\alpha$ emission, the plasma volume integrated $n_e$ and the $T_e$ and He II ($n=4\rightarrow3$) emissivity at mid-radius, are shown on the left of Fig. \ref{fig:trace}. The duration of the $I_p$ flat-top and diffusion time is of the order of 100 ms and the global energy confinement time is of the order of 10 ms. The plasma is therefore sufficiently stationary during the transport analysis. The absence of ELMs in L-mode allows for the study of neoclassical and turbulent transport without the perturbations induced by intermittent MHD phenomena. However, on MAST, magnetohydrodynamic (MHD) activity is always present from a certain time onwards during the discharge due to the evolution of the safety factor profile, $q$, leading either to the onset of sawteeth activity or to an internal $m=1$, $n=1$ kink instability known as the `long-lived mode' (LLM) \cite{Chapman2010}. This MHD activity introduces additional transport whose effects are difficult to disentangle from those of neoclassical and turbulent transport. We reduce the transport analysis time window to a few tens of milliseconds to avoid sawteeth activity (see Fig. \ref{fig:trace}), however the LLM occurs near the end of the transport analysis time window ($t\sim0.28$ s) and is unavoidable. 

Various diagnostics were used to obtain the background plasma parameters. The Thomson scattering diagnostic \cite{Scannell2008} provided measurements of $n_e$ and $T_e$, while the ion temperature, $T_i$, and toroidal velocity, $v_{\phi}$, measurements were obtained from CX measurements of carbon \cite{Conway2006}. $Z_{eff}$ profiles are based on the He$^{2+}$ and C$^{6+}$ density measurements made by the RGB diagnostic \cite{Patel2004b} (described in the next subsection). The \textsc{efit++} code \cite{Appel2006} is used to provide a magnetic flux reconstruction, with the poloidal magnetic field constrained by the pitch angle measured by the motional Stark effect (MSE) diagnostic \cite{DeBock2008} and the boundary of the plasma constrained by the edge $D_{\alpha}$ emission. Time averaged profiles of $T_{e,i}$, $n_e$, $v_{\phi}$, $q$ and $Z_{eff}$ are plotted on the right of Fig. \ref{fig:trace} as a function of $\rho$, where $\rho$ is the square root of the normalised toroidal flux. Lastly, the 2D beam emission spectroscopy (BES) diagnostic \cite{Dunai2010} was positioned to measure turbulent electron density fluctuations near the plasma edge with a radial and poloidal resolution of $\sim$2 cm \cite{GhimKim2010}; measurements from this diagnostic will be analysed in section \ref{sec:measurements}.

From the reference scenario described above we performed a two-point $I_p$ scan at fixed $B_T$ by lowering $I_p$ from 900 kA to 600 kA. To compensate for the lower confinement at lower $I_p$ and keep the plasma temperature similar to the 900 kA scenario, we also increased the additional heating power from 2.1 MW to 3.2 MW. By comparing the time averaged background plasma profiles on the right of Fig. \ref{fig:trace} for the 900 kA (green lines) and 600 kA (blue lines) L-mode plasmas, the resulting target plasma had a higher edge safety factor, $q_{95}$, with respect to the reference scenario (6.3 instead of 5.5); density and temperature profiles on the other hand were very similar to the reference scenario. $Z_{eff}$ is $\sim1$ in the core of both L-mode plasmas, with a moderate increase found at 600 kA. Again, sawteeth have been been avoided in the 600 kA L-mode plasma, although the LLM begins near the end of the analysis time window at $t\sim0.25 $ s. 

Finally, we performed a confinement scan by inducing a controlled L-H transition during the reference plasma at $t=0.25$ s. This was done by lowering the vertical position of the magnetic axis, originally 5 cm above the vessel equator, by $\sim3$ cm, which on MAST is known to induce a prompt transition to an ELMy H-mode as shown by the red $D_{\alpha}$ trace in Fig. \ref{fig:trace}. Due to the difference in confinement, the density and (to a lesser extent) temperature profiles were different as shown in Fig. \ref{fig:trace}. $Z_{eff}$ is close to unity across the entire plasma radius in H-mode. Furthermore, the LLM is unavoidable throughout the H-mode period, which causes a decrease in $v_{\phi}$ within the $q=1$ surface. We therefore only analyse the H-mode scenario beyond the $q=1$ surface ($\rho \geq 0.4$). 

\subsection{Helium Spectroscopy}
\begin{figure}[t]
	\centering
		\includegraphics[width=0.90\textwidth]{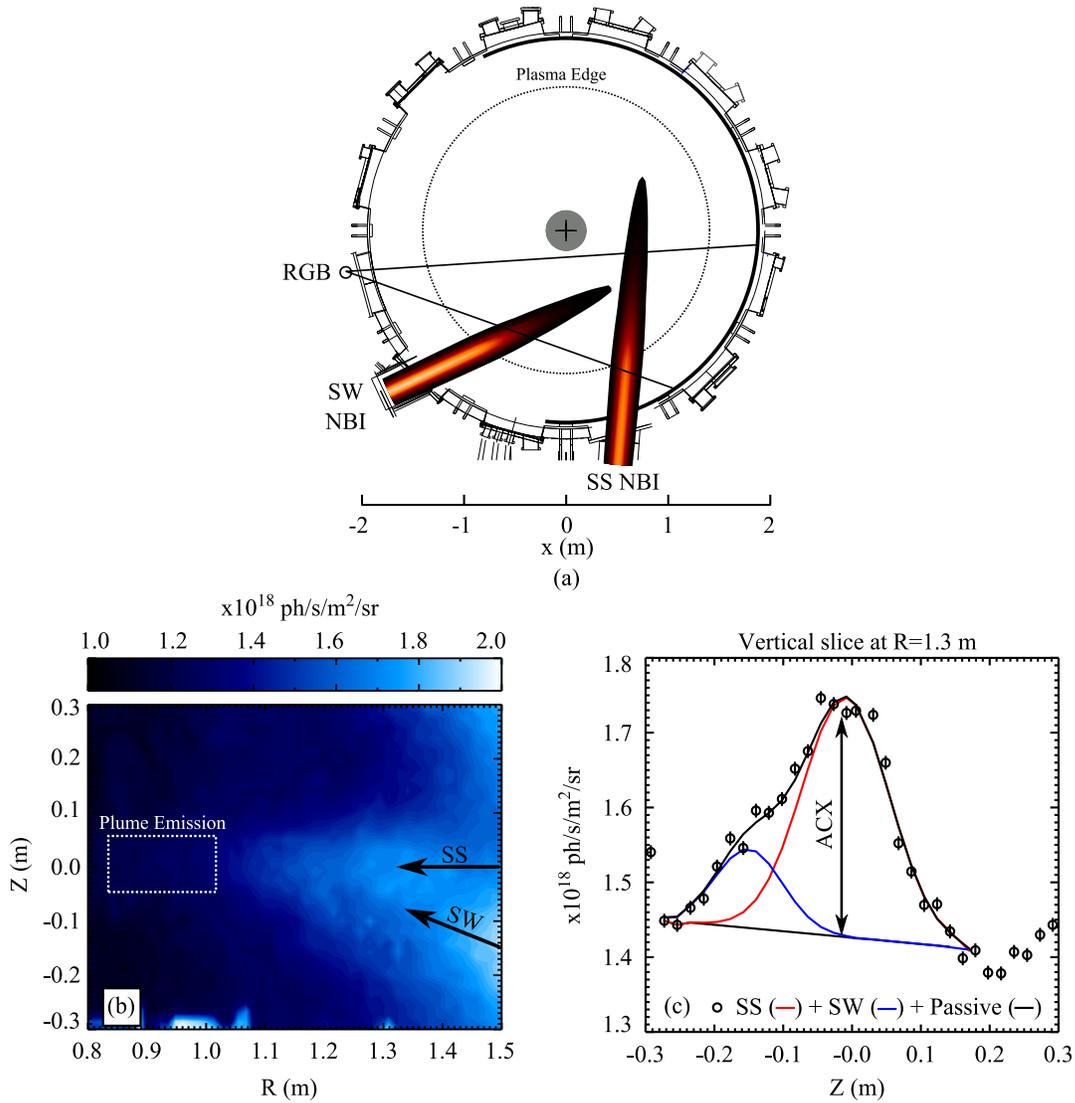}
	\caption{The plan view of MAST in (a) illustrates the horizontal field of view used in (b). Modelled SS and SW neutral beam density contours are shown for reference. He II ($n=4\rightarrow3$) emissivity contours, measured at $t=0.22$ s during the L-mode 600 kA plasma (\#29417) using the RGB spectral band-pass ($\lambda=468.86$ nm; $\Delta\lambda\sim5$ nm), are illustrated in (b); $R$ and $Z$ are with respect to the SS beam axis. The vertical Gaussian fits at $R=1.3$ m for the SS and SW ACX contributions are indicated by the red and blue lines respectively in (c).}
	\label{fig:he_cxrs}
\end{figure}

Helium gas was puffed into the vessel using a piezo valve located on the inboard side above the lower divertor. To describe the influx of helium into the plasma, the piezo valve was calibrated using a screened Bayard-Alpert fast ionisation gauge. At a plenum pressure of 1.5 Bar, it was found that 25 ms gas puffs, injecting a total of $\sim10^{19}$ He atoms, were sufficient for a transport analysis. The injected helium increases the He$^{2+}$ concentration from $\sim$4 \% to $\sim$8 \% (as shown later in Fig. \ref{fig:intdens}) and can be considered a tracer. The timing of the puff, illustrated at the bottom of Fig. \ref{fig:trace}, was chosen at the beginning of the $I_p$ flat-top for the L-mode scenarios and at the beginning of the H-mode period, which allowed for at least $\sim50$ ms of transport analysis in each case. The transport analysis time window is smallest in H-mode due to the onset of sawteeth at $t\sim0.32$ s. 

An imaging diagnostic on MAST, called RGB \cite{Patel2004b}, is located $\sim$20 cm above the equator and views the full plasma cross-section using a video graphics array (VGA) sensor giving a spatial resolution along the SS beam axis of $\sim$3 mm at a frame rate of 200 Hz. To average out random noise, every 2 frames are averaged and each 2D frame having 640 x 480 pixels is rebinned over 5 pixels (128 x 96 pixels) giving a reduced temporal and radial resolution of 100 Hz and 1.5 cm respectively. Chord-integrated emissivities from six different spectral band-passes in the red, green and blue regions of the visible spectrum are all available through one viewing iris. The $\sim$5 nm spectral band-pass centred on $\lambda=468.86$ nm (referred to as the blue channel) measures the He II ($n=4\rightarrow3$) active charge exchange (ACX) spectrum line at 468.5 nm induced by neutral beam atoms. 

The horizontal field of view (FOV) measured by RGB is illustrated in Fig. \ref{fig:he_cxrs}a. Notice that the FOV measures ACX emission from both (SS and SW) beams. The beams lie along the machine equator, therefore the $\sim20$ cm vertical elevation of RGB allows for the separation of both ACX components, as shown in Fig. \ref{fig:he_cxrs}b and c. Both beams are operating in the L-mode 600 kA plasma, while only the SS beam is operating in the L-mode and H-mode 900 kA plasmas. For consistency, only ACX emission from the SS beam is considered in the three plasmas. 

A single 2D frame of emission measured through the blue channel during the L-mode 600 kA plasma is illustrated in Fig. \ref{fig:he_cxrs}b; height ($Z$) and radius ($R$) are given with respect to the SS beam axis. A passive CX feature (PCX) induced by thermal deuterium neutrals and a low temperature electron impact component is also measured through each pixel and must be subtracted; here we simply refer to both features as the PCX component. PCX emission along the beam axis is estimated using a linear fit of the emissivity above and below the beam, demonstrated by the black line in Fig. \ref{fig:he_cxrs}c. The validity of this assumption was tested using a plasma where no beams were operating. A single or double Gaussian fit (depending on the the number of beams operating) is applied to the remaining emission, where the Gaussian peak is taken as the ACX component. An example of the fitting technique at $R=1.3$ m is illustrated by the red and blue solid lines in Fig. \ref{fig:he_cxrs}c. This procedure is applied along each pixel (or $R$) column to obtain a radial ACX emissivity profile. 

After the ACX process is complete, the donated electron decays to the ground state of the He$^+$ ion and then becomes either ionised or re-excited into the $n=4$ shell by electron impact. The resultant He II ($n=4\rightarrow3$) spectral line can significantly contribute to non-local active sight lines as the He$^+$ ions traverse along the magnetic field line \cite{Fonck1984a}; at the point of creation of the ground state He$^+$ ion, the electron excitation emission is typically an order of magnitude less than the ACX emission. These secondary emitting ions are known as plume ions. Since RGB is located close to the machine equator, plume ions created on a field line with a relatively large pitch angle will be directed away from the mid-plane and therefore do not contribute. On the other hand, a plume ion originating from regions of the plasma with moderately low pitch angle will travel almost horizontally and contribute to the non-local ACX emissivities. 

The lowest values of pitch angle are typically found within $\rho \leq 0.4$ (corresponding to $R\leq1.2$ m). Toroidal ion speeds in this region are $\geq 100$ km/s, allowing the plume ions to travel approximately one toroidal revolution before ionisation. A weak cloud of emission near the magnetic axis, thought to be emission from the plume ions created in the region $R \leq 1.2$ m, can be seen in Fig. \ref{fig:he_cxrs}b. ACX emissivities, and hence He$^{2+}$ density profiles, are therefore only quoted from $\rho \geq 0.2$. 

\section{Helium Density Measurements}\label{sec:measurements}

\begin{figure}[t]
	\centering
		\includegraphics[width=0.55\textwidth]{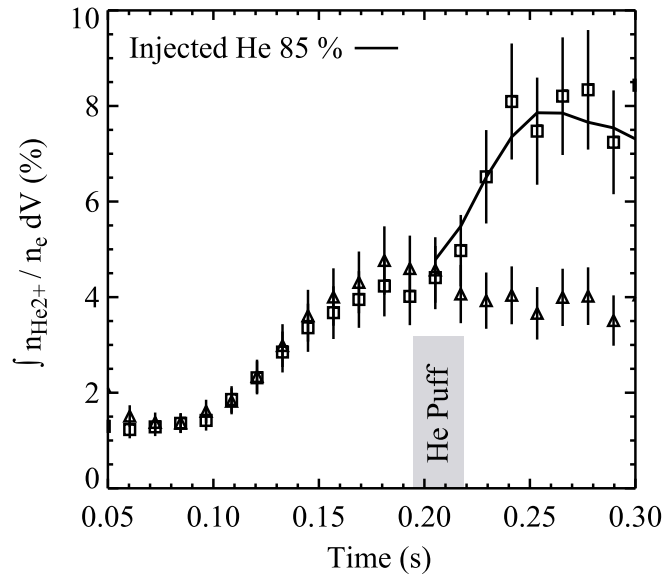}
	\caption{Symbols illustrate the He$^{2+}$ concentration integrated over the plasma volume for the L-mode 900 kA scenario (squares $-$ puff \#29261 and triangles $-$ no puff \#29424).}
	\label{fig:intdens}
\end{figure}

Chord-integrated measurements of the ACX emission, $\epsilon_{He^+}$, described in the previous section are converted into local $n_{He2+}$ measurements using the expression in Eq. \ref{Hedens}, where $q_{CX}$ is the effective CX emission coefficient interpolated from the Atomic Data Analysis Structure (ADAS) \cite{Summers2004} and $\int{n_b^F ds}$ is the modelled chord-integrated density of each neutral beam fraction, $F$.
\begin{equation}\label{Hedens}
n_{He2+}=\frac{4\pi \int{\epsilon_{He^+}ds}}{\sum_{F}q_{CX} \int{n_b^Fds}}
\end{equation}
Simulations of the neutral beam density were carried out on MAST using a code based on the narrow beam approximation method \cite{Feng1995,Schneider2011}. A source of systematic error may stem from the secondary source of donor electrons associated with thermal beam `halos' \cite{Synakowski1990}. Inclusion of an artificial halo, simulated using an additional thermal beam species fraction with a density and $n=2$ principal quantum electron shell population equal to the main beam fraction, decreases $n_{He2+}$ by $< 3$ \% and is thus considered negligible.

To observe the temporal evolution due to the injected helium gas, the background $n_{He2+}$ profiles are calculated for plasma scenarios with no gas puff and subtracted. Figure \ref{fig:intdens} gives an example of the helium concentration, $n_{He2+}/n_e$, integrated over the plasma volume during both puff and no puff discharges along with the total number of injected helium neutrals from the gas puff. Since the background $n_{He2+}$ concentration is well matched before the gas puff, the fuelling efficiency can be approximated from these curves to be $\sim$0.85. Similar fuelling efficiencies can be found for both the 600 kA L-mode and 900 kA H-mode plasmas. 

Functional fits are applied to $n_{He2+}$ to average out the noise. It has been found by Wade et al. \cite{Wade1995} that a good representation of $n_{He2+}$, which is smooth in time and space, can be made by firstly fitting the following temporal functions to each spatial location as
\begin{equation}\label{wadefit}
n_{He2+}(\rho,t) = 
\begin{cases} 
                    A_1+A_2\tanh[A_3(t-A_0)]       & \mbox{$\rho$} \leq 0.4\mbox{; t} > 0 \\ 
                    A_1+A_2\exp[-A_3(t-A_0)^2]     & \mbox{$\rho$} > 0.4 \mbox{; t} \leq A_0   \\
										A_1+A_2\exp[-A_4(t-A_0)^2]     & \mbox{$\rho$} > 0.4 \mbox{; t} > A_0 
\end{cases}
\end{equation}
Once values of $A_1$, $A_2$, $A_3$ and $A_4$ have been determined for every spatial point, a $5^{th}$ order polynomial fit is applied to the spatial profile for every moment in time. Carrying out the fitting in this order produced smooth gradients in both time and space; a necessary requirement in determining the transport coefficients as discussed in section \ref{sec:transport}. 

\begin{figure}[p]
	\centering
		\includegraphics[width=0.95\textwidth]{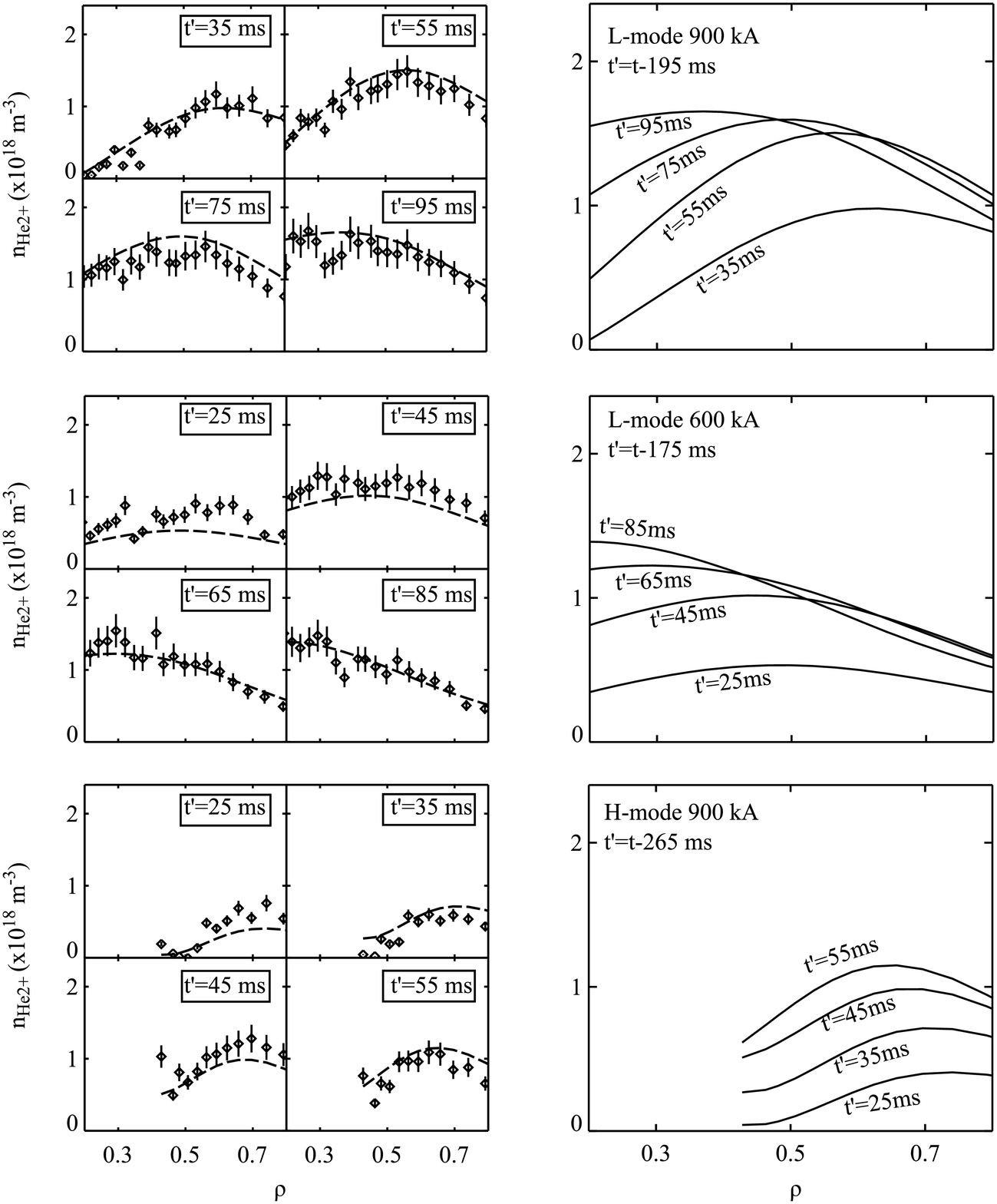}
	\caption{On the left, the graphs compare the fitted and experimental $n_{He2+}$ profiles at four different times after the gas puff for each scenario. The four fitted lines are shown together in separate plots on the right to demonstrate the evolution of the $n_{He2+}$ profile.}
	\label{fig:polyfit}
\end{figure}

The spatial evolution of the fitted data in each plasma scenario are compared with the experimental data points in Fig. \ref{fig:polyfit}. A clear perturbation of $n_{He2+}$ is observed over the range $0.2 \leq \rho \leq 0.8$ during L-mode at both high and low $I_p$ and  $0.4 \leq \rho \leq 0.8$ in H-mode. The He$^{2+}$ ions reach the core in a faster time scale at low $I_p$ suggesting a higher radial diffusivity and inward convection. Analysis of the time window before the onset of sawteeth in H-mode suggests that the He$^{2+}$ ions stagnate around $\rho\sim0.6$. 

\subsection{Experimental Transport Model}\label{sec:transport}
To evaluate the transport of the He$^{2+}$ ions, either a predictive or an interpretative method can be used. The predictive method relies on an impurity transport code that solves the transport equation,
\begin{equation}\label{continuity}
\frac{\partial n_{He2+}}{\partial t}=-\nabla \cdot \Gamma_{He2+}+Q.
\end{equation}
where $Q$ represents the source and sink terms which couple He$^{2+}$ to the neighbouring ionisation stages. The He$^{2+}$ flux, $\Gamma_{He2+}$, is described using diffusion, $D_{He}$, and convection velocity, $v_{He}$, coefficients as 
\begin{equation}\label{flux}
\Gamma_{He2+}=-D_{He}\nabla n_{He2+}+v_{He}n_{He2+}.
\end{equation}
The zero flux peaking factor can therefore be defined as 
\begin{equation}\label{zerofluxpeak}
\frac{1}{L_{n_{He2+}}}=-\frac{v_{He}}{D_{He}}
\end{equation}
where $\frac{1}{L_{\alpha}}=-\nabla \alpha/\alpha$ with $\alpha$ representing any plasma profile. The peaking factor is used later to compare the derived $-v_{He}/D_{He}$ ratio with $1/L_{n_{He2+}}$ calculated from the reference plasma with no gas puff and also with the quasilinear peaking factor calculated from the gyrokinetic analysis.

Density profiles of each ionisation stage are calculated by \textsc{sanco} \cite{Lauro1994} by numerically solving Eq. \ref{continuity}. Inputs of temperature and density mapped to magnetic flux surfaces allow \textsc{sanco} to calculate the source and sink terms based on rates supplied by ADAS for electron impact ionisation from lower ionisation states and electron recombination from higher charge states (edge parameters such as the recycling rate, $\eta$, and the parallel loss time, $\tau_{||}$, along with the temporal evolution of the neutral influx are also required as inputs). The least-squares fitting code, \textsc{utc} \cite{Whitefordthesis,Giroud2007}, is used to determine radial profiles of $D_{He}$ and $v_{He}$ that produce the best match between experimental and simulated $n_{He2+}$ profiles. 
\begin{figure}[t]
	\centering
		\includegraphics[width=0.95\textwidth]{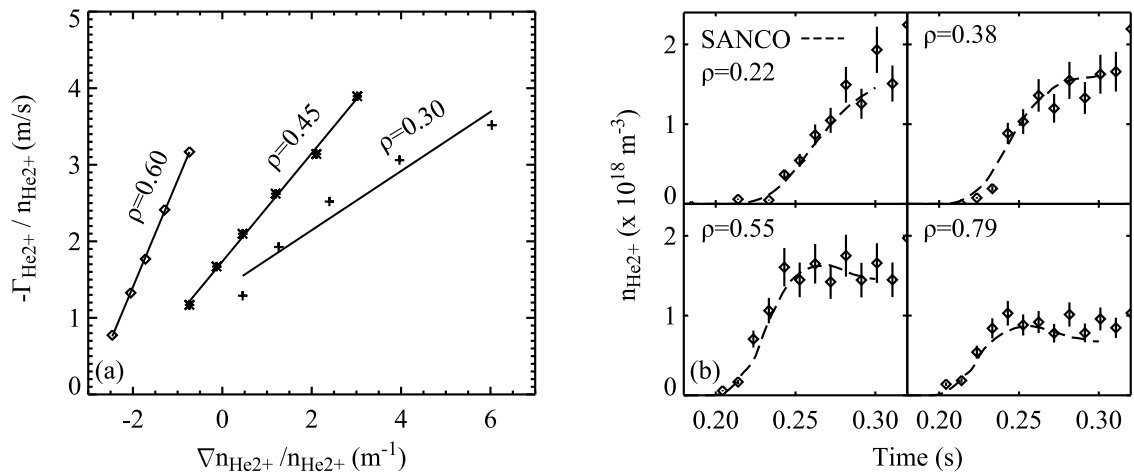}
	\caption{This figure illustrates an example of the fitting procedures required to determine the He transport coefficients for the L-mode 900 kA plasma. An example of the linear relationship between the helium particle flux and spatial gradient is demonstrated in (a). At each $\rho$, the different points represent the temporal evolution (from right to left) of the flux and gradient. The experimental and simulated \textsc{sanco} $n_{He2+}$ profiles (b) are shown as a function of time at four different values of $\rho$.}
	\label{fig:transport_evaluation}
\end{figure}

The interpretative approach involves integrating Eq. \ref{continuity} assuming $Q=0$ to give
\begin{equation}\label{invertcontinuity}
\left . \frac{1}{A(\rho)n_{He2+}}\frac{\partial N_{He2+}}{\partial t}\right|_0^{\rho}=D_{He}\frac{\nabla n_{He2+}}{n_{He2+}}-v_{He}
\end{equation}    
where $A$ and $N_{He2+}$ are the area and volume integrated helium density respectively inside each magnetic flux surface $\rho$. If $D_{He}$ and $v_{He}$ are only functions of $\rho$ and not time, then the temporal evolution of the term on the left hand side of Eq. \ref{invertcontinuity}, which is $-\Gamma_{He2+}/n_{He2+}$, is a linear function of $\nabla n_{He2+}/n_{He2+}$ with the gradient and offset at each radial location representative of $D_{He}$ and $-v_{He}$ respectively. The obvious advantage of this method is the fact it is more deterministic than predictive in nature. Furthermore it does not rely on any predetermined values of edge parameters. However, this method breaks down and $D_{He}$ and $v_{He}$ cannot be determined in regions of the plasma where $Q$ becomes significant. This interpretative approach will be referred to as the Flux Gradient (FG) method. In an attempt to provide a robust evaluation of the transport coefficients, both the UTC-SANCO and FG methods are compared for each scenario. 

\begin{figure}[t]
	\centering
		\includegraphics[width=0.95\textwidth]{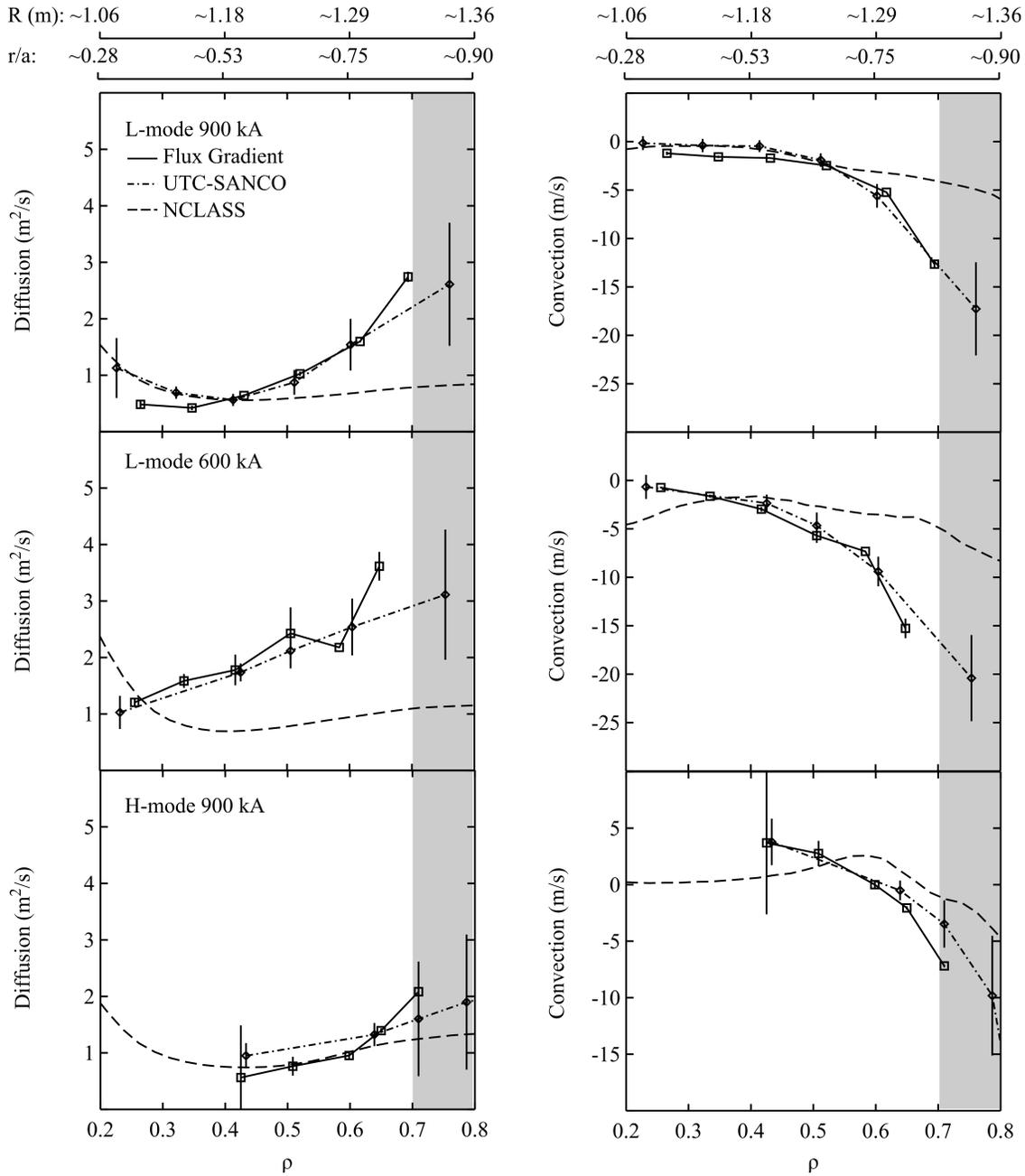}
	\caption{Helium transport coefficients, evaluated using the Flux Gradient (FG) method (squares) and the UTC-SANCO method (diamonds), are plotted for each plasma scenario. The neoclassical rates (single dash) are calculated by \textsc{nclass}, while the shaded region represents the region of significant sources for the FG method.}
	\label{fig:transport}
\end{figure}

An example of the fitting procedures involved in both the FG and UTC-SANCO method are shown for the 900 kA L-mode scenario in Fig. \ref{fig:transport_evaluation}. The temporal evolution of $-\Gamma_{He2+}/n_{He2+}$ as a function of $\nabla n_{He2+}/n_{He2+}$ during the density rise after the He gas puff is shown in Fig. \ref{fig:transport_evaluation}a for $\rho=[0.3,0.5,0.7]$. The different time points of each flux surface are described by a linear fit, as shown by the solid lines in Fig. \ref{fig:transport_evaluation}a. For the UTC-SANCO method, estimates must be made of the edge parameters described previously. A helium glow discharge is run prior to every discharge on MAST, therefore helium is assumed to have a high recycling rate of $\eta\sim0.95$; varying $\eta$ by $\pm0.05$ didn't produce any significant change to the transport coefficients outside of the error bars. $\tau_{||}$ can be estimated using $Rq_{95}/v_{th}$ with a thermal energy of $\sim$1 eV, a plasma edge location of $R=1.4$ m and $q_{95}=5.8$ which gives $\tau_{||}\sim$1 ms. The neutral helium influx profile can be estimated from the calibrations of the impurity gas valve along with the determined fuelling efficiency of 0.4. Comparisons of the UTC-SANCO and experimental $n_{He2+}$ are shown in Fig. \ref{fig:transport_evaluation}b. 

\begin{figure}[t]
	\centering
		\includegraphics[width=0.55\textwidth]{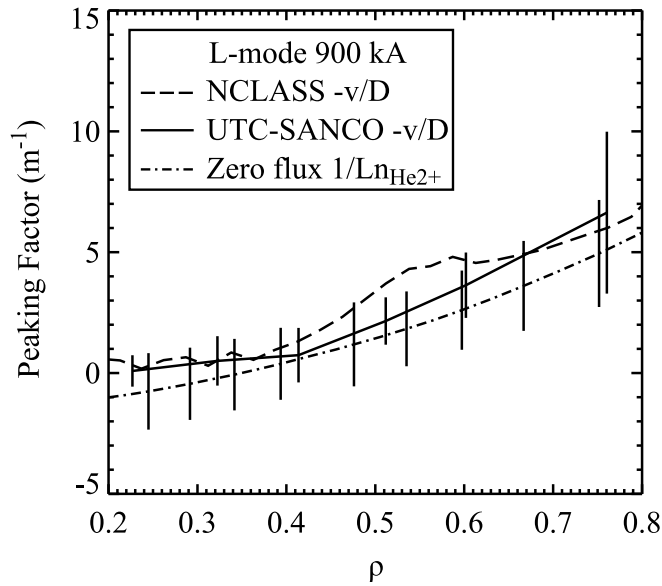}
	\caption{Peaking factor for the L-mode 900 kA plasma calculated using the $D_{He}$ and $v_{He}$ profiles from UTC-SANCO analysis (solid line), the zero flux peaking factor from the plasma without a gas puff (dash-dot line) and the neoclassical $D_{He}$ and $v_{He}$ coefficients from \textsc{nclass}.}
	\label{fig:peakingfactor}
\end{figure}

Evaluated diffusivity, $D^{exp}_{He}$, and convective velocity, $v^{exp}_{He}$, coefficients are shown as a function of $\rho$ for both the FG and UTC-SANCO method in Fig. \ref{fig:transport}. Agreement within error bars is found between both transport models in each plasma scenario. A general trend for each scenario is an increase of $D^{exp}_{He}$ with radius from $\rho=0.2$ up to $\rho=0.8$ and a peaked inward pinch (denoted by $v^{exp}_{He}<0$) in the edge region of the plasma. For the L-mode plasma at 900 kA, $D^{exp}_{He}$ ranges from 0.5 m$^2$/s in the core to $<$ 3 m$^2$/s near the plasma edge, while $v_{He}^{exp}\sim-15$ m/s is observed near the plasma edge. An example of the ratio $-v/D$ for this plasma is compared with the zero flux peaking factor, $1/L_{n_{He2+}}$, from the equivalent plasma with no gas puff in Fig. \ref{fig:peakingfactor}. Agreement within error bars is found between the two peaking factors, which is expected since the assumption is that the injected helium does not modify the transport coefficients.

Decreasing $I_p$ in L-mode causes a moderate increase in $D_{He}^{exp}$ and $v_{He}^{exp}$ over the range $0.3\leq\rho\leq0.6$.  For the confinement scan at 900 kA, a relatively small decrease in $D_{He}^{exp}$ is found over the range $0.5\leq\rho\leq0.8$ in H-mode compared to the L-mode plasma. The pinch found near the edge of the L-mode plasma is also decreased in the H-mode plasma. Over the range $0.4<\rho<0.6$, there is evidence that the convection changes direction from inwards to outwards. This outward convection, coupled with the inward convection near the plasma edge, is likely causing the moderate build up of $n_{He2+}$ around $\rho=0.65$.

In past helium transport studies on conventional tokamaks, the ratio of $D_{He}^{exp}/\chi_T^{eff}$ is often examined to indicate whether a certain plasma scenario will be viable for a DT plasma producing He ash \cite{Synakowski1990,Synakowski1993,Wade1995} and therefore it is important to briefly discuss this ratio in this paper for the purpose of expanding these results to STs. We define the total effective heat diffusivity, $\chi_T^{eff}$, as
\begin{equation}\label{effheat}
\chi_T^{eff}=\frac{Q_e+Q_i}{n_e\nabla T_e+n_i\nabla T_i}
\end{equation}
where $Q_{e,i}$ is the electron and ion heat fluxes respectively calculated from a local power balance analysis using the predictive transport code \textsc{jetto} \cite{Cenacchi1988,Cenacchi1988a}. The time averaged profiles of $n_e$, $T_{e,i}$, $q$ and $Z_{eff}$ are required as inputs for \textsc{jetto}.

\begin{figure}[t]
	\centering
		\includegraphics[width=0.55\textwidth]{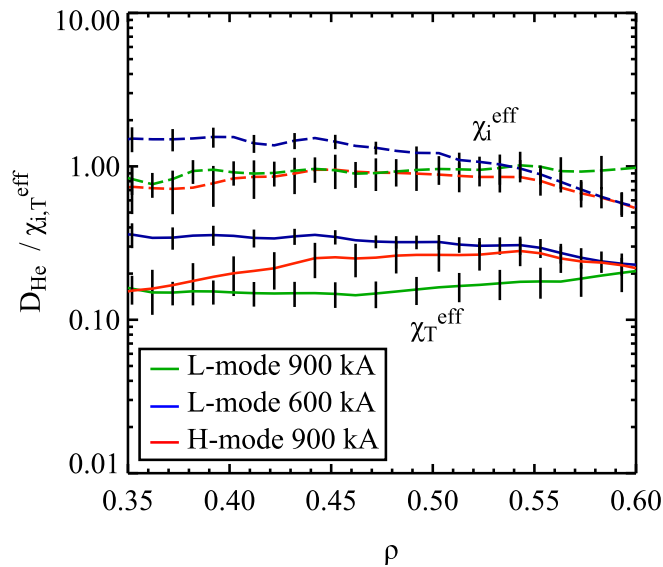}
	\caption{The magnitude of $D_{He}/\chi_{T,i}^{eff}$ is shown for each plasma in the radial range where the $T_e$ and $T_i$ gradients are well defined.}
	\label{fig:chi}
\end{figure}
The ratio of $D_{He}^{exp}/\chi_T^{eff}$ is shown in Fig. \ref{fig:chi} for the range where $\nabla T_{e,i}$ are well defined. In each plasma scenario, the magnitude of $D_{He}^{exp}/\chi_T^{eff}$ is $\sim0.25$, however closer agreement in magnitude is found between $\chi_i^{eff}$ and $D_{He}^{exp}$ with $D_{He}^{exp}/\chi_i^{eff}\sim1$. Generally on MAST, $\chi_e$ is predominantly driven by turbulence in the electron channel while $\chi_i$ is of the same order of magnitude as neoclassical predictions \cite{Roach2009}. The latter conclusion is in agreement with the previous helium transport studies mentioned in the previous paragraph which indicate a ratio of $D_{He}^{exp}/\chi_i^{eff}\sim1$. A previous study on MAST states that the ratio of $D_D/\chi_T^{eff}\sim0.4$, suggesting that the confinement time of He$^{2+}$ ions is similar to the main ions \cite{Valovic2005}. 
 
\subsection{Role of Neoclassical Transport}\label{sec:neo}

\begin{figure}[t]
	\centering
		\includegraphics[width=0.55\textwidth]{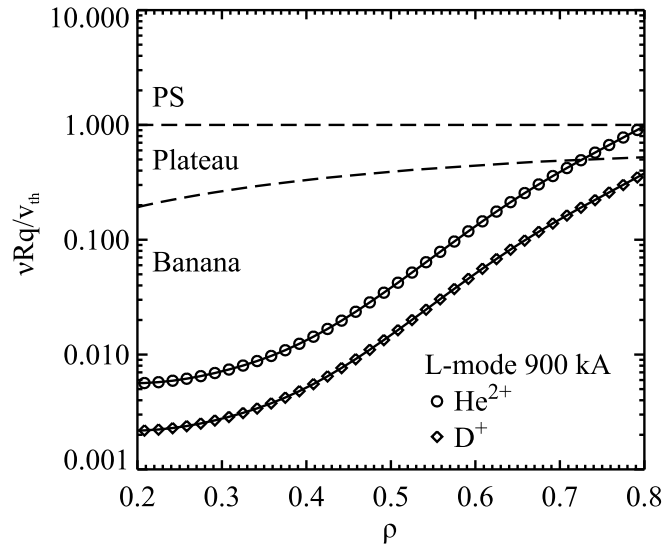}
	\caption{Variation of the He$^{2+}$ and main ion collisionality with $\rho$ for the 900 kA L-mode plasma. Dashed lines represent the cut-offs for the Banana, Plateau and Pfirsch-Schl\"{u}ter collisionality regimes.}
	\label{fig:neo_regime}
\end{figure}

Neoclassical diffusivity $D^{NC}$ and convection $v^{NC}$ coefficients have been obtained from the \textsc{nclass} code \cite{Houlberg1997} within \textsc{jetto} and shown by the dashed lines in Fig. \ref{fig:transport}. The \textsc{nclass} simulations use the time averaged profiles of $n_e$, $T_{e,i}$, $q$ and $Z_{eff}$ shown in Fig. \ref{fig:trace}, while the magnetic flux reconstruction was carried out with \textsc{efit++}, as described in section \ref{sec:scenarios}. For the typical densities and temperatures measured in each plasma scenario, the He$^{2+}$ ions are well within the Banana-Plateau (BP) regime, as shown in Fig. \ref{fig:neo_regime} by the calculated He$^{2+}$ collisionality, $\nu_{He2+}^*=\nu_i Rq/v_{th}$, where $\nu_i$ is the ion-ion collision frequency. 

In the BP regime, $D^{NC}$ can be shown \cite{Guirlet2006a} to depend mainly on the background plasma profiles as
\begin{equation}\label{bp_diffusion}
D^{NC} \propto \frac{q^2 m_i n_i} {\epsilon^{3/2} B_T^2 \sqrt{T_i}}
\end{equation}   
Here $\epsilon$ is the inverse aspect ratio and $m_i$, $T_i$, and $n_i$ are the main ion mass, temperature and density respectively. $v^{NC}$ can also be written as
\begin{equation}\label{bp_convection}
v^{NC}=-D^{NC}\left( z \frac{1}{L_{n_D}} + H^{BP} \frac{1}{L_{T_i}}\right)
\end{equation}
where $z$ is the impurity charge and the factor $H^{BP}$ is related to the viscosity of the impurity and main ions and is typically negative, meaning that the inward $T_i$ and $n_i$ gradients typically drive an outward and inward impurity flux respectively. 

In L-mode, the experimental transport coefficients are significantly greater than the neoclassical predictions in the range $0.5<\rho\leq0.8$ and $0.3<\rho\leq0.8$ for the 900 kA and 600 kA plasmas respectively. For the H-mode plasma, the region of anomalous transport is smaller and confined to the range $\rho>0.7$. There is little difference between $D^{NC}$ in all three plasma scenarios. The main changes in $v^{NC}$ for each plasma scenario are caused mainly by the differences in the $n_e$ gradient. In the range $\rho<0.3$, an inward $v^{NC}$ is found at 600 kA but not at 900 kA in the L-mode plasmas. This is due to the steep $n_e$ gradient found in this range in at 600 kA. The effect of the $n_e$ gradient is also apparent in the H-mode plasma. In the range $0.5\leq\rho\leq0.7$, $v^{NC}$ is directed outwards due to the outward $n_e$ gradient found in this range. The flatter $n_e$ gradient found near the H-mode plasma edge causes the neoclassical pinch to decrease compared to the L-mode case. 

The ratio of $-v^{NC}/D^{NC}$ for the 900 kA L-mode plasma is in good agreement with the UTC-SANCO $-v^{exp}/D_{He}^{exp}$ ratio over the range $0.2\leq\rho\leq0.8$, as shown in Fig. \ref{fig:peakingfactor}. The situation is also similar for the L-mode and H-mode plasmas. An analysis of the zero flux peaking factors alone therefore suggests that the transport is neoclassical over the entire analysed radial range. On the other hand, the time-dependent analysis shows that the transport is anomalous near the plasma edge in L-mode. This illustrates the importance of performing time-dependent impurity experiments to determine the absolute magnitude of the $D_{He}^{exp}$ and $v_{He}^{exp}$ profiles. 

\subsection{Role of Anomalous Transport}\label{sec:gyro}

\begin{table}[bp]
\caption{Background plasma parameters for $\rho=0.7$} 
\centering 
\begin{tabular}{|c|c|c|c|} 
\hline 
  & L-mode 900 kA & L-mode 600 kA & H-mode 900 kA \\ [0.5ex] 
\hline 
$c_s$ (m/s)          & 1.3x10$^5$      & 1.1x10$^5$      & 1.6x10$^5$      \\   
$a$ (m)              & 0.47            & 0.46            & 0.48 \\
$q$                  & 2.00            & 2.40            & 2.21 \\ 
$\hat{s}$            & 3.63            & 3.82            & 4.75 \\
$\nu^*_e$            & 1.45            & 2.18            & 1.89 \\
$a/L_{T_e}$          & 7.42            & 6.70            & 4.74 \\
$a/L_{T_i}$          & 4.30            & 4.01            & 4.31 \\
$a/L_{n_e}$          & 1.89            & 2.09            & 0.06 \\
$u'$                 & 0.75            & 0.87            & 0.85 \\
$\gamma_E$           & 0.18            & 0.19            & 0.09 \\
\hline 
\end{tabular}
\label{table:gs2param} 
\end{table}

Linear gyrokinetic simulations yield the properties of the most unstable microinstabilities. Strictly, more demanding non-linear gyrokinetic simulations are required to predict the properties of the saturated turbulence. In this paper we use a quasilinear approach, exploiting the properties of the dominant linear modes to model the saturated turbulence. We therefore determine the dominant electromagnetic microinstabilities as a function of normalised wave number $k_y\rho_i$ (where $\rho_i$ is the ion Larmor radius) using two local flux-tube gyrokinetic codes, \textsc{gs2} \cite{Kotschenreuther1995,Dorland2000} and \textsc{gkw} \cite{Peeters2009}. This will reveal the stability of microinstabilites driven by the ion temperature gradient (ITG) at long wavelength ($k_y\rho_i < 1$), electron temperature gradient (ETG) at short wavelength ($k_y\rho_i > 3$) and the trapped electron modes (TEM) at mid-wavelength ($1 \leq k_y\rho_i \leq 3$). Then we will compare the quasilinear helium transport coefficients associated with the dominant linear modes, with the impurity transport coefficients measured in experiment.  

\begin{figure}[t]
	\centering
		\includegraphics[width=0.9\textwidth]{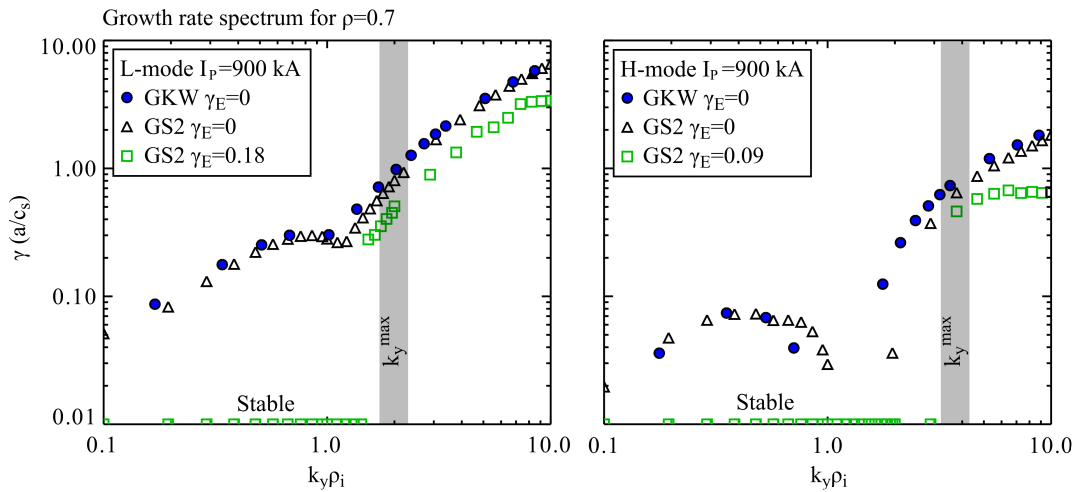}
	\caption{Linear growth rate spectrum for flux surface $\rho=0.7$ in L-mode (a) and H-mode (b) plasmas at $I_p=900$ kA. Calculations were performed with and without flow shear using the \textsc{gs2} (squares,triangles) and \textsc{gkw} (circles) codes respectively. The dashed lines represent the dominant linear mode in each case. Symbols on the x-axis represent the stable modes found with $\gamma=0.18$.}
	\label{fig:gs2growth}
\end{figure}

The flux surface $\rho=0.7$ was chosen for analysis since this surface provides reliable experimental transport coefficients dominated by turbulence. Table \ref{table:gs2param} lists the relevant plasma equilibrium parameters of each scenario at $\rho=0.7$. These parameters include the sound speed $c_s=\sqrt{T_i/m_i}$, the minor radius $a$, the safety factor $q$, magnetic shear $\hat{s}=(r/q)$ d$q/$d$r$, electron collisionality $\nu^*_e=\nu_ea/c_s$, plasma beta $\beta$, the gradients $a/L_{T_i}$, $a/L_{T_e}$ and $a/L_{n_e}$ and the equilibrium flow shear $\gamma_E=u'/q$, where $u'$ is the spatial gradient of the parallel ion rotation velocity normalised by $-a/c_s$. Note that in each simulation we assume that $n_i=n_e$ and set helium as trace. On spherical tokamaks like MAST, the stabilising perpendicular component, $\gamma_E$, of the sheared toroidal flow, induced by strong neutral beam injection, dominates over the destabilising parallel component, and stabilises long wavelength modes driven by ITGs \cite{Roach2005}. Growth rates of the dominant instabilities have been computed in the absence of sheared flow with both codes. We have also obtained effective linear growth rates, $\gamma$, that account for the stabilising influence of flow shear, from \textsc{gs2} simulations using the method described in \cite{Roach2009}.

The growth rate spectrum from the microstability analysis of the L-mode and H-mode discharges at 900 kA using \textsc{gs2} and \textsc{gkw} are shown in Figs. \ref{fig:gs2growth}a and b respectively for the range $0.1 < k_y\rho_i < 10$. For L-mode, both codes show unstable modes across the $k_y\rho_i$ spectrum with $\gamma_E=0$. A similar growth rate spectrum is observed in H-mode with $\gamma_E=0$, however modes in the region $1 < k_y\rho_i < 2$ are stable. A similar stable region of the growth rate spectrum was reported for an earlier MAST H-mode discharge in \cite{Roach2009}. The main difference between the local equilibria in L- and H-mode is that $a/L_{n_e}$ is lower in H-mode, making collisionless TEMs more stable \cite{Dannert2005}. The sensitivity of TEM stability to the density gradient, was recently demonstrated in the microstability analysis of MAST discharges with pellet injection \cite{Garzotti2014}. 

\begin{figure}[t]
	\centering
		\includegraphics[width=0.55\textwidth]{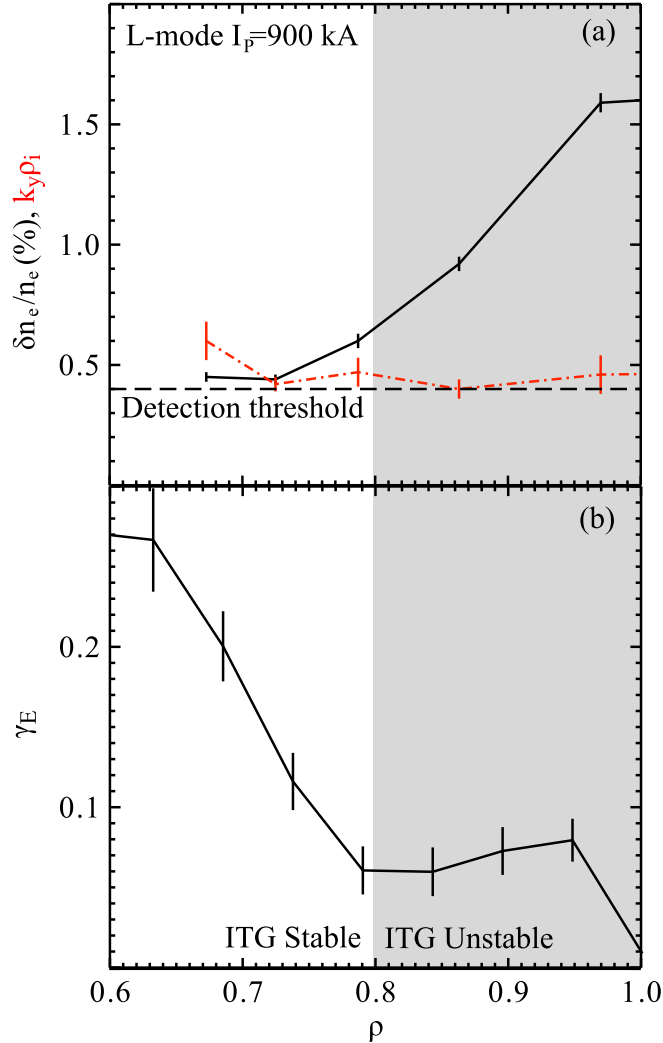}
	\caption{The BES electron density fluctuation signal with the corresponding $k_y\rho_i$ (dashed dot line) (a) and the time averaged equilibrium flow shear (b) are shown for the L-mode plasma at $I_p=900$ kA.}
	\label{fig:bes}
\end{figure}

With $\gamma_E$ set to the experimental value in the simulation, the plots in Figs. \ref{fig:gs2growth}a and b suggest that unstable long wavelength ITG modes become suppressed in L- and H-mode at $\rho=0.7$. The maximum mixing length diffusivity, $\sim \gamma/k_y^2$, arises at $k_y^{max}$ equal to $k_y\rho_i$ $\sim2$ and $\sim4$ for L- and H-mode respectively. The BES electron density fluctuation signal, $\delta n_e/n_e$, averaged over 4 poloidal channels at each radius and time averaged over 2 ms ($\pm$ 1 ms either side of 0.24s), is plotted in Fig. \ref{fig:bes}a. The values of $k_y\rho_i$ observed by the BES diagnostic correspond to ITG turbulence. For $\rho=0.7$, the electron density fluctuation signal is approximately 0.4 \%, very close to the detection threshold. This suggests that ITG modes are indeed stable for this flux surface, but does not prove or disprove the existence of TEMs. Towards the edge of the plasma the BES signal increases indicating a rising amplitude of ITG turbulence. This could be explained by the fact that the equilibrium flow shear decreases with radius as shown in Fig. \ref{fig:bes}b. Furthermore, $\chi_i^{eff}$ is close to neoclassical in this discharge, as is typical on MAST \cite{Valovic2011} because ITG modes are often stabilised by sheared flow \cite{Roach2005, Roach2009}. 

\begin{figure}[t]
	\centering
		\includegraphics[width=0.55\textwidth]{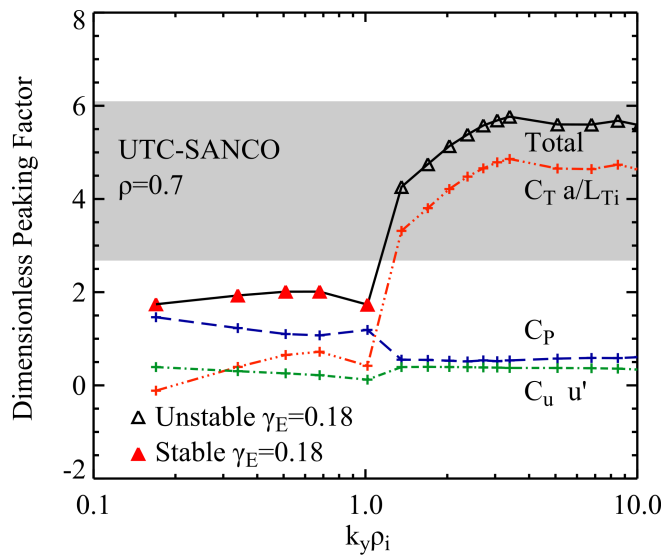}
	\caption{Gyrokinetic analysis of the peaking factor using the \textsc{gkw} code is shown as a function of $k_y\rho_i$ for the L-mode 900 kA plasma at $\rho=0.7$ assuming $\gamma_E=0$. }
	\label{fig:gkwflux}
\end{figure}

We now exploit quasilinear theory to obtain the helium transport coefficients associated with the dominant linear modes by using the expression for the particle flux as \cite{Angioni2009}
\begin{equation}
\label{gyroflux}
\frac{a\Gamma^{GKW}_{He2+}} {n_{He2+}} =D_{He}^{GKW} \left( \frac{a}{L_{n_{He2+}}} + C_T\frac{a}{L_{T_{i}}} + C_uu' + C_p \right)
\end{equation}
The thermodiffusive ($C_T$) \cite{Coppi1978}, rotodiffusive ($C_u$) \cite{Camenen2009} and convective ($C_p$) \cite{Angioni2006} dimensionless coefficients depend on the impurity mass, charge and temperature as well as the background turbulence. We keep the same notation as in Eq. \ref{continuity} where a negative particle flux represents an inward motion of impurities. In recent studies (see for example \cite{Fulop2009,Skyman2014}), the dimensionless zero flux peaking factor, $a/L_{n_{He2+}}=-av_{He}/D_{He}$, calculated from gyrokinetic simulations in the limit of $\Gamma_{He2+}=0$, is of value in describing steady state density profiles that can be achieved in the absence of a central source. In this analysis, the dimensionless peaking factor is simply used to compare with experiment to determine whether agreement occurs in the TEM region. 

The UTC-SANCO ratio $-v_{He}^{exp}/D_{He}^{exp}$ at $\rho=0.7$ is interpolated from Fig. \ref{fig:peakingfactor}; note that the normalisation factor used in \textsc{gkw} to remove the m$^{-1}$ dependence of this ratio is $a=0.86$, which is different to \textsc{gs2} with $a=0.47$. Gyrokinetic simulations, assuming $n_e=n_i$ and with the He impurity as a trace species, can then be used to determine quasilinear estimates of the zero flux peaking factor as a function of $k_y\rho_i$ using 
\begin{equation}
\label{peak}
\frac{a}{L_{n_{He2+}}}=-(C_T\frac{a}{L_{T_i}}+C_uu'+C_p)
\end{equation}
The impurity flux transport coefficients $C_T$, $C_u$ and $C_p$ are determined as a function of $k_y \rho_i$ using \textsc{gkw} by calculating the particle flux of four trace amounts of helium with a pre-determined orthogonal set of equilibrium gradients \cite{casson2013}. Substituting $C_T$, $C_u$, $C_p$ and $a/L_{T_i}$ and $u'$ (measured by CX) into Eq. \ref{peak} provides the zero flux peaking factor as a function of $k_y\rho_i$. An estimate for the overall zero flux peaking factor is finally obtained by assuming that impurity transport is dominated by modes at $k_y = k_y^{\rm max}$ where the mixing length diffusivity $\gamma/k_y^2$ is maximised. 

Fig. \ref{fig:gkwflux} illustrates the quasilinear zero flux peaking factor for the L-mode 900 kA plasma calculated with $\gamma_E=0$. The red filled triangles represent the points that can be assumed to be zero if flow shear is included. Encouraging agreement with experiment is found in both direction and amplitude of the zero flux peaking factor around $k_y^{max}$. The contribution from each of the different linear components of the flux are shown by the dashed lines. The major contribution to the peaking factor comes from $a/L_{T_i}$, with $C_T$ essentially only determining the direction. The $C_u$ and $C_p$ coefficients are also directed inwards but only contribute weakly to the peaking factor. 

Quasilinear estimates of the diffusion and convection coefficients, for the 900 kA L-mode discharge at $\rho=0.7$, can be made by using the expression 
\begin{equation}\label{quasilinearflux}
\Gamma_{He2+}^{QL}=\Gamma_{He2+}^{GKW}\frac{\chi_e^{eff}}{\chi_e^{GKW}}
\end{equation}
where $\chi_e^{GKW}=Q_e^{GKW}T_e/(a/L_{T_e})$. We interpolate the value of $\chi_e^{eff}$ at $\rho=0.7$ from the radial profiles shown in Fig. \ref{fig:chi} and choose the value of $\chi_e^{GKW}$ situated around $k_y^{max}$. This gives values of $D_{He}^{QL}$ of the order of $1-10$ m$^2$/s, similar to experimental diffusivities. Furthermore, a previous study by Casson et al. \cite{casson2013} showed that this quasilinear method of obtaining of the peaking factor gives good agreement with full non-linear gyrokinetic simulations.  


\section{Discussion and Conclusions}\label{sec:conclusions}
This paper provides a robust evaluation of the fully ionised helium transport on MAST made from 2D visible charge exchange measurements taken during a two point $I_p$ scan in L-mode and a comparison of L- and H- mode at constant plasma current. During the $I_p$ scan, a decrease in both diffusion and inward convection is found at high $I_p$. In H-mode, the magnitude of the inward convection near the plasma edge decreases, while the convection at mid-radius changes to an outward direction. The helium particle and heat diffusivity ratio is $D_{He}$/$\chi_T^{eff}\sim 0.25$ and $D_{He}/\chi_i^{eff}\sim1$, a result similar to conventional tokamaks. The L-mode plasmas are dominated by anomalous transport from mid-radius to the plasma edge region; the H-mode plasma on the other hand is dominated by neoclassical transport. 

Anomalous transport observed in L-mode on the flux surface $\rho=0.7$ has been analysed with \textsc{gs2} and \textsc{gkw}. A number of results suggest that TEM turbulence is driving the helium transport. Firstly, simulations suggest that ITG turbulence is stabilised by equilibrium flow shear at $\rho=0.7$, which is in agreement with BES measurements that find electron density fluctuations at ITG wavenumbers that are minimal on this surface, but more significant towards the edge. Secondly, the quasilinear zero flux peaking factor for helium, calculated by \textsc{gkw} at $\rho=0.7$, agrees with experiment in both magnitude and direction in the TEM region. The main contribution to the magnitude and direction of the peaking factor comes from the ion temperature gradient and the thermodiffusion coefficient respectively. Lastly, the most significant difference between the microinstability growth rate spectrum in L- and H-mode is the stabilisation of the TEMs in H-mode due to the lower electron density gradient.

From the three plasma scenarios discussed, the results indicate that an ELMy H-mode would be the most favourable scenario to transport helium ash out of the plasma core of future fusion STs due to the outward convection found in the H-mode plasma core. The electron density gradient plays a crucial role in the transport of light impurities in spherical tokamaks; a steep electron density gradient causes an inward pinch of helium directly from neoclassical transport and from turbulence associated with collisionless TEMs. Generally the helium pinch caused by turbulence dominates over the neoclassical pinch in L-mode. Since the neoclassical convection caused by the electron density gradient is amplified by the impurity charge and impurities with different charge respond to different scale length microinstabilities, an assessment of higher charge impurities, carbon and nitrogen, during the same plasma scenarios is now underway. 

\section*{Acknowledgements}
This work was partly funded by the RCUK Energy Programme grant number EP/I501045 and the European Communities under the contract of Association between EURATOM and CCFE. To obtain further information on the data and models underlying this paper please contact PublicationsManager@ccfe.ac.uk. The views and opinions expressed herein do not necessarily reflect those of the European Commission. The gyrokinetic calculations were carried out on HECToR supercomputer (EPSRC grant EP/H002081/1).

\section*{References}
\bibliographystyle{prsty}
\bibliography{paper_bib}
\end{document}